\documentclass{aa}
\usepackage{graphicx}
\usepackage{xcolor}

\newcommand{\beq}{\begin{equation}}
\newcommand{\beqa}{\begin{eqnarray}}
\newcommand{\eeq}{\end{equation}}
\newcommand{\eeqa}{\end{eqnarray}}

\def\vec#1{\ensuremath{\mathchoice{\mbox{\boldmath$\displaystyle#1$}}
{\mbox{\boldmath$\textstyle#1$}}
{\mbox{\boldmath$\scriptstyle#1$}}
{\mbox{\boldmath$\scriptscriptstyle#1$}}
}}
\begin{document}
\title{Launching of asymmetric outflows from the star-disk magnetosphere}
\titlerunning{Launching of asymmetric outflows \ldots}
\authorrunning{\v{C}emelji\'{c} et al.}
\author{M. \v{C}emelji\'{c},
\inst{1,2,3,4}
A. Kotek,
\inst{1}
\and
W. Klu\'{z}niak
\inst{1}
 }
 \institute{Nicolaus Copernicus Astronomical Center of the Polish Academy of Sciences, Bartycka 18, 00-716
Warsaw, Poland, \email{miki@camk.edu.pl}
\and
Nicolaus Copernicus Superior School, College of Astronomy and Natural Sciences, Gregorkiewicza 3, 87-100, Toru\'{n}, Poland
\and
 Research Centre for Computational Physics and Data Processing, Institute of Physics, Silesian University in Opava, Bezru\v{c}ovo n\'am.~13, CZ-746\,01 Opava, Czech Republic
\and
Academia Sinica, Institute of Astronomy and Astrophysics, P.O. Box 23-141,
Taipei 106, Taiwan
}
\date{Received ??; accepted ??}
\abstract
{In resistive and viscous magnetohydrodynamical simulations, we obtain axial outflows launched from the innermost magnetosphere of a star-disk system. The launched outflows are found to be asymmetric. We find the part of the parameter space corresponding to quasi-stationary axial outflows and compute the mass load and angular momentum flux in such outflows. We display the obtained geometry of the solutions and measure the speed of propagation and rotation of the obtained axial outflows. 
}


\maketitle

\section{Introduction}
In spite of over a century of investigation, astrophysical jets, which are highly collimated and energetic outflows of matter, are still a very active field of research at all scales. Observations of jets, collected at steadily increasing resolution, are now reaching the innermost regions of the active galactic nuclei, black hole X-ray binaries, young stellar objects (YSOs), symbiotic objects with white dwarfs and compact binaries with neutron stars, but the detailed launching mechanism is still not known in any of the cases. Although the difference in scales is huge, similarities between the jets indicate possible common mechanism, but for any such claim, more work is needed for each particular class of objects. Understanding of jet launching and propagation is a crucial input for the description of evolution of objects at all scales, as it determines the mixing of material in the interstellar as well as intergalactic scales. 

One-sided outflows with complex magnetic fields were studied in \cite{LovRom10,RomRev14}. We find asymmetric launching of outflows with dipole stellar magnetic field. In \cite{cem19} we devised a setup for simulations of star-disk magnetospheric interaction, based on \cite{ZF09,ZF13}. We confirmed their results, which corroborated findings in \cite{R09}. In a parameter study for the thin accretion discs around slowly rotating stars, we studied stars rotating up to 20\% of the Keplerian rotational rate at the stellar equator \citep{cembrun23}. We also investigated midplane backflow in such discs \citep{MishraBackfl23} and initially probed the effects of geometries of the stellar field  different from dipolar \citep{Ciec22}. 

Here we extend our simulations to the faster rotating stars, in which we obtain axial outflows of different mass loads and velocities in opposite directions, above and below the disk midplane. By presenting different axial outflows from one and the same kind of object such results  provide information about the influence of local environment and magnetic field strength and geometry on the launching of outflows. Such axial outflows would play an important role in the large scale jet launching mechanism.

In the following Section~\ref{set} we describe our numerical setup, with overview of the results in Section~\ref{res} and conclusions in Section~\ref{con}.

\section{Numerical setup}\label{set}
We perform 2D axisymmetric star-disk simulations with the publicly available PLUTO code \citep{m07} and setup as in \cite{cem19}, only that now we compute in a complete $\theta\in [0,\pi]$ co-latitudinal plane. This way we do not prescribe the disk mid-plane as a boundary condition, but evolve the equations in the complete disk self-consistently. Here we work with the resolution $R\times\theta=[125\times 100]$ grid cells in a physical domain reaching the radius of 50 stellar radii, $R\in [1,50]R_\star$, with a logarithmically stretched radial grid in spherical coordinates $(R,\theta,\varphi)$. We chose the larger, almost double radial extension of the physical domain than was used in \cite{cem19}, to ensure a larger stability of the flow and less influence of the outer radial boundary. The overall resolution is lower than in the previous simulations, but with our choice of the logarithmically stretched radial grid, the innermost part of the magnetosphere is still well resolved. We checked that the results are qualitatively similar to the corresponding simulations in the larger resolution.   

The disk is set up following \cite{Kita95,KK00}, with the addition of a hydrostatic, initially non-rotating corona \citep{cemklupar23} above the rotating star and a dipole moment of the stellar field parallel with the stellar rotation axis. The viscosity and resistivity are parameterized by the Shakura-Sunyaev prescription, with the anomalous viscous and resistive dissipative coefficients $\alpha_{\rm v}$ and $\alpha_{\rm m}$.  Much larger than their microscopic counterparts, such diffusive coefficients are free parameters in the simulation. They are motivated by the assumption that dissipation is a result of macroscopic instabilities, i.e., turbulence \citep{SS73,BH91}. A condition for inclusion of the diffusive term is defined by $\beta=P_{\rm mag}/P_{\rm hyd}>0.5$, meaning that the magnetic pressure is prevailing. Where this inequality is not satisfied, the diffusive terms are set to zero. For maintaining the divergence-less magnetic field we use the constrained transport method. The magnetic field we set using the split-field approach, in which only changes from the initial stellar magnetic field are evolved in time \citep{Tan94,Pow99}. We use the second-order piecewise linear reconstruction, with a Van Leer limiter in density and a \texttt{minmod} limiter in pressure and velocity, RK2 time stepping, and modified Roe solver. The solver modification is such that flags in the \texttt{flag\_shock} subroutine do not switch in the presence of shocks, except when the internal energy becomes lower than 1\% of the total energy. In our setup, all the viscous heating is locally radiated away from the disk. We use the polytropic equation of state with the plasma polytropic index $\gamma=5/3$. To avoid issues with the midplane backflow in the disc \citep{MishraBackfl23}, we perform simulations with the anomalous viscosity coefficient $\alpha_{\rm v}=1$, in which there is no such backflow.

\section{Asymmetric axial outflows}
\label{res}
\begin{figure}
\includegraphics[width=1\columnwidth]{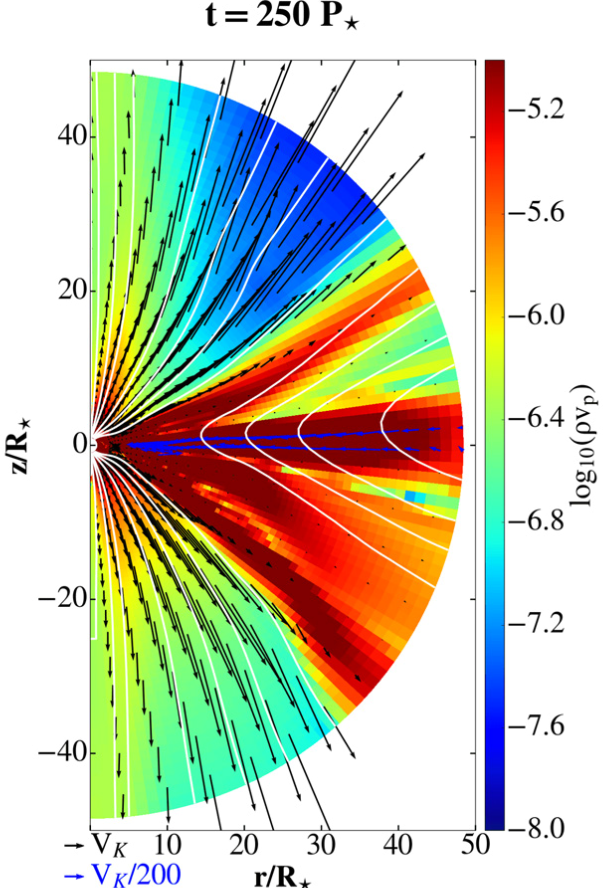}
\caption{Momentum flux in a logarithmic color grading in a snapshot at t=250~$P_\star$ from our simulation with the stellar magnetic field B$_\star$=250 G, stellar rotation rate $\Omega=0.5\Omega_{\rm br}$ and the resistivity coefficient $\alpha_{\rm m}=0.4$. Normalization of the velocity vectors in the disk (blue arrows) is two hundred times the velocity in the magnetosphere outside of the disk (black arrows), as the velocity in the disk is much smaller. With white solid lines is shown a sample of the magnetic field lines. Time is measured in the number of stellar rotations at the equator, P$_\star$, and momentum is measured in the code units.}
\label{jet250}
\end{figure}
\begin{figure*}
\includegraphics[width=0.93\columnwidth]{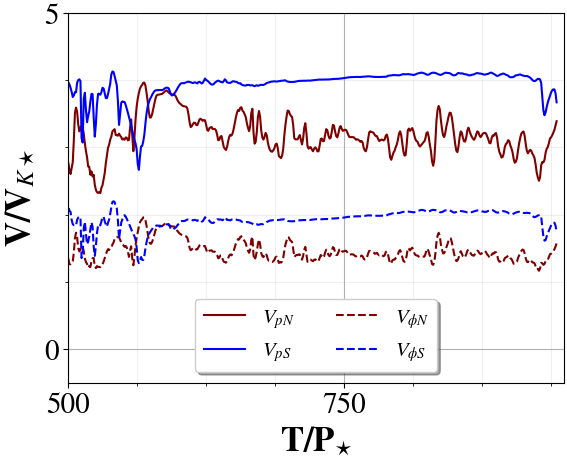}
\includegraphics[width=0.99\columnwidth]{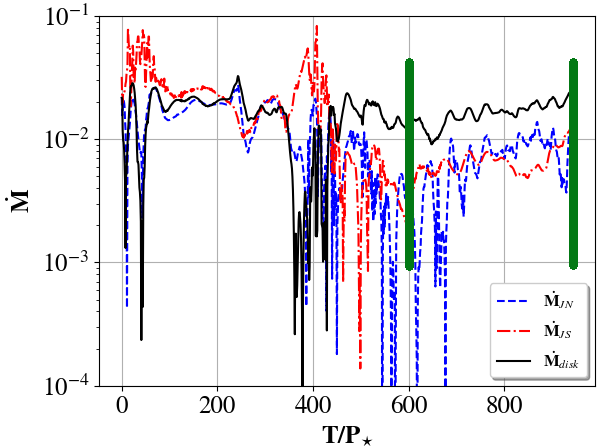}
\caption{Analysis of the result from Fig.~\ref{jet250}. All the values are evaluated at the distance 25R$\star$. {\it Left panel}: poloidal ($V_{\rm p}$) and azimuthal (V$_\varphi$) components of the axial outflow velocities, in the units of Keplerian velocity at the stellar equator. Velocities of the outflow above the disk midplane are shown with blue, and of the outflow below the disk midplane with the red lines. {\it Right panel}: mass flux in the outflows and disk, during the whole simulation, in the units of $10^{-7} M$/y. With the black solid line is shown the mass flux through the disk, and with blue and red dashed lines through the outflows above and below the disk midplane, respectively. The interval when the quasi-stationary state is reached is marked with the vertical green thick solid lines. Figure adopted from \cite{Kotek20}.}
\label{figb}
\end{figure*}
\begin{figure*}
\includegraphics[width=0.99\columnwidth]{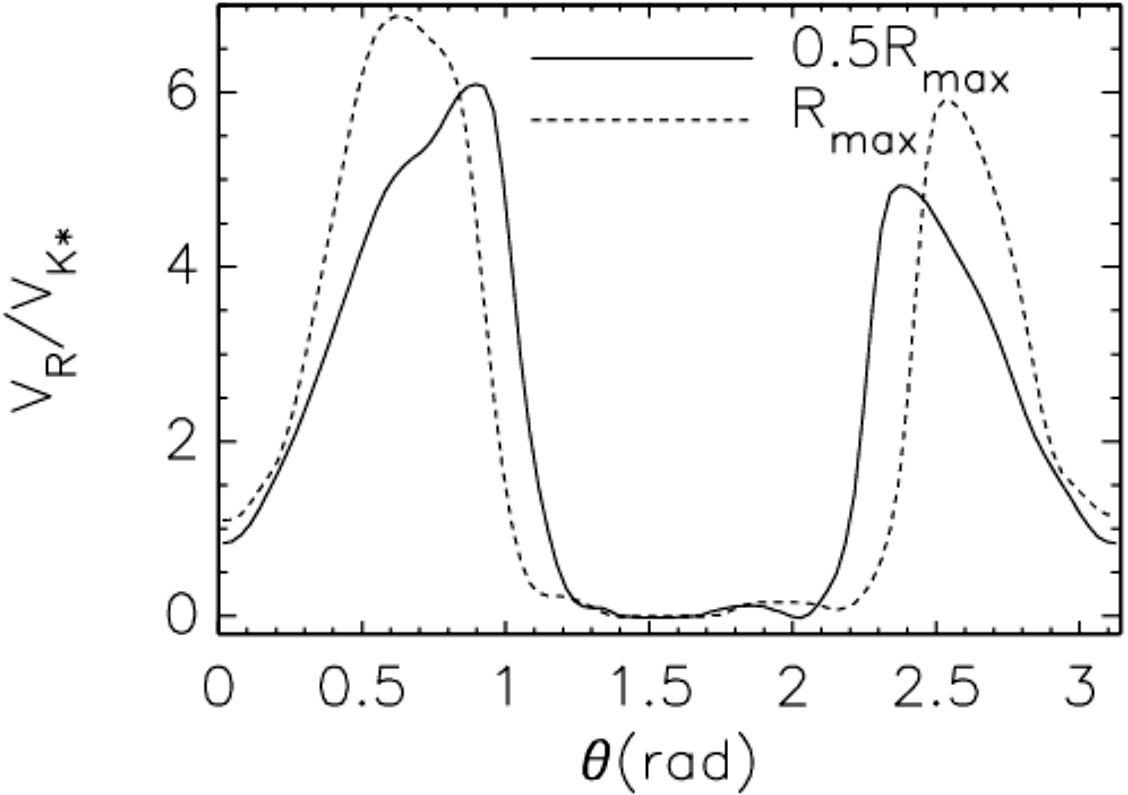}
\includegraphics[width=0.99\columnwidth]{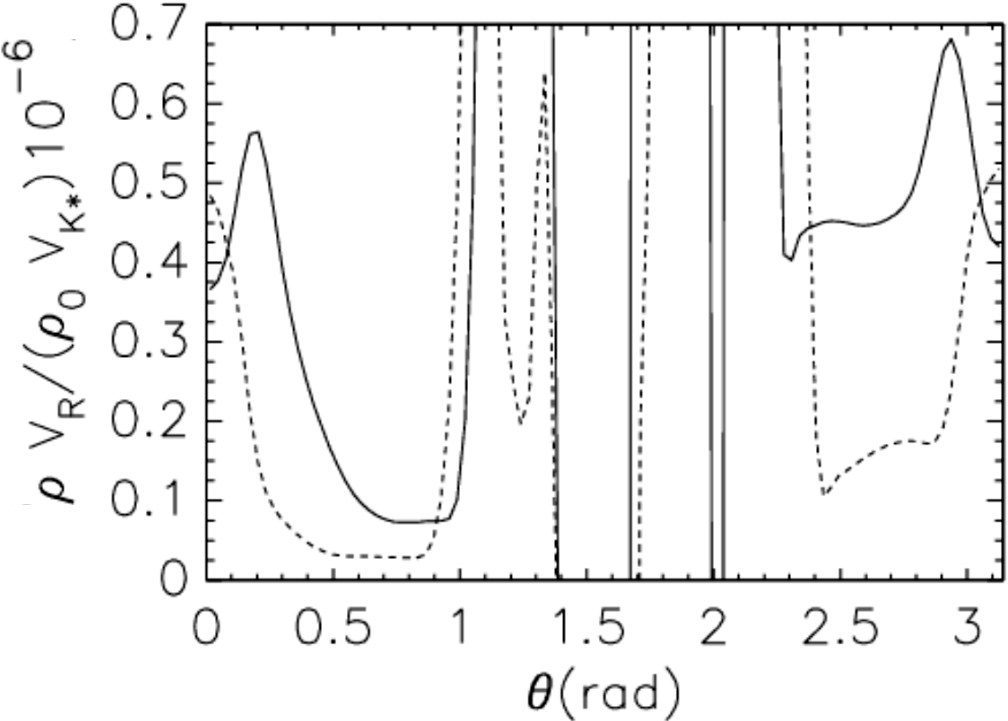}
\caption{{\it Left panel}: radial propagation velocity of the outflow shown in Fig.~\ref{jet250}. {\it Right panel}: mass flux $\rho v_{\rm r}$ in the same quasi-stationary solution. Inside the disk, the mass flux is much larger so for $\theta=1-2$ in the given range we capture only the bottom of the profiles. The axial outflows are positioned at about $\theta=0.3$ and $\theta=2.7$ radians. The density is given in units of $\rho_0=10^{-10}g~cm^{-3}$. In both panels, the velocities are shown in units of the Keplerian velocity at the stellar equator and computed along the half-circles in the meridional plane at $R=25R_\star$ and at $R=50R_\star$.}
\label{figc}
\end{figure*}
\begin{figure*}
\includegraphics[width=2\columnwidth]{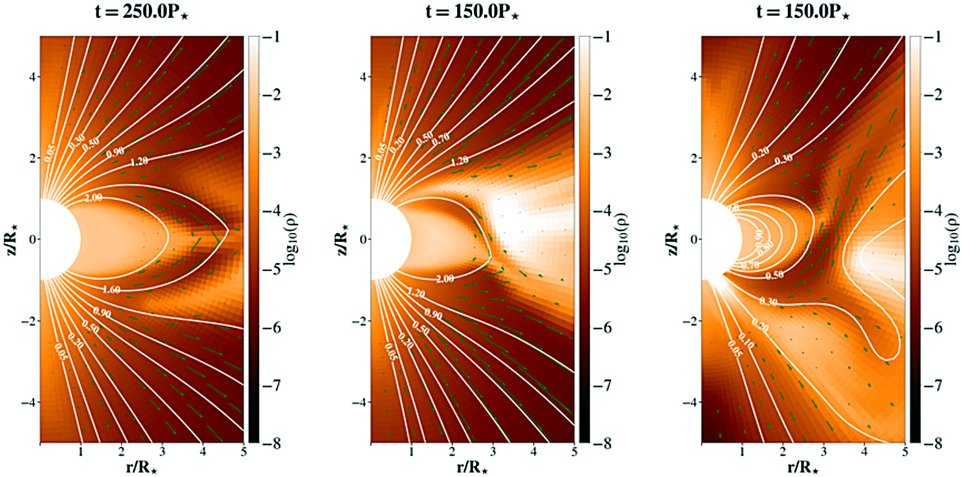}
\caption{{\it Left}, {\it middle} and {\it right} panels show the density distribution in the DA, DCA and DCEA geometries, respectively. These plots are obtained with the choice of parameters  $\Omega_\star/\Omega_{\rm br}=(0.5, 0.8, 0.5)$, B$_\star$ = (1000, 1000, 250) G, $\alpha_{\rm m}$ = (1, 0.4, 0.1) for the (left, middle and right) panels, respectively.}
\label{troj}
\end{figure*}
\begin{figure}
\includegraphics[width=\columnwidth]{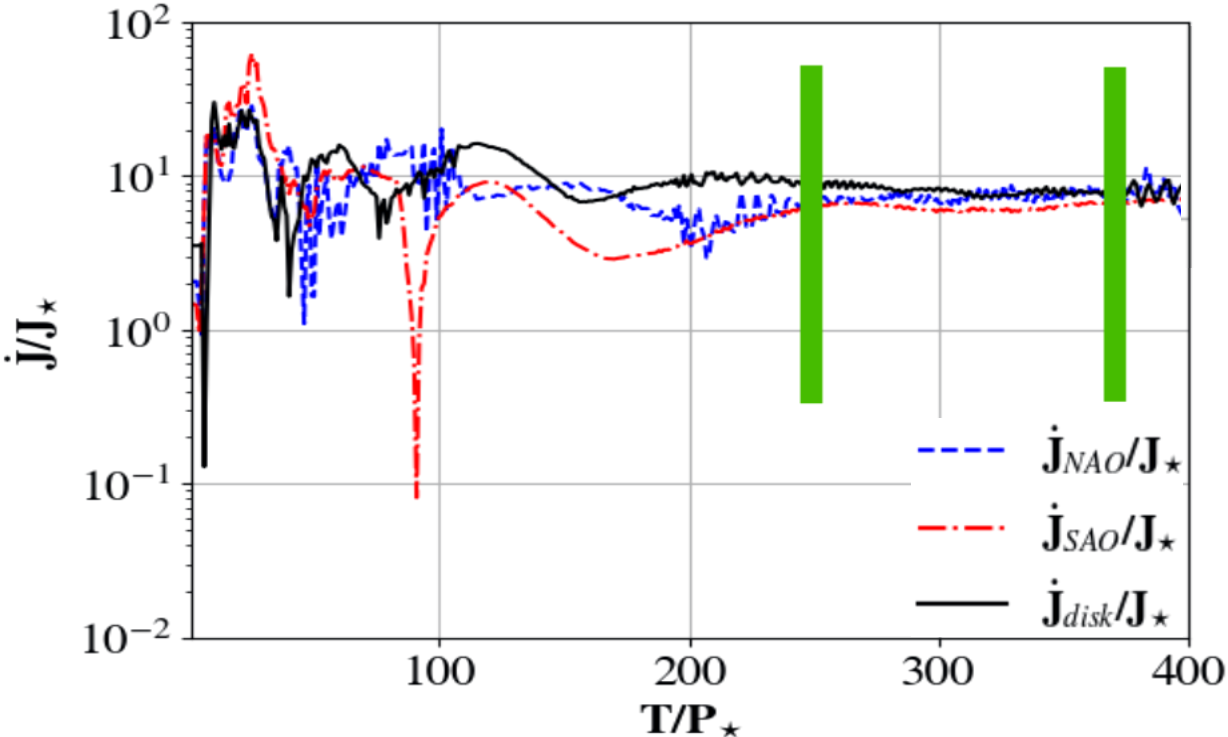}
\caption{Angular momentum flux $\dot{J}$ in the simulation with $\Omega_\star=0.8\Omega_{\rm br}$, B$_\star=500$~G, $\alpha_{\rm m}=0.4$. The blue curve corresponds to North Axial Outflow (NAO), the red curve corresponds to South Axial Outflow (SAO) and the black line corresponds to the disk contribution. The thick vertical green lines define the time interval over which the simulation is quasi-stationary. Such an interval was used to define the average value of $\dot{J}$ in Table~\ref{jdottab} in the text.}
\label{jall}
\end{figure}
We report quasi-stationary solutions in our simulations, with asymmetric axial outflows launched from the magnetosphere of a star-disk system in opposite directions \citep{Kotek20, Kotek22}. 

The mass and angular momentum fluxes are obtained by integrating
\begin{eqnarray}
\dot{M}=\int_{\rm S}\rho\vec{\rm v}_{\rm p}\cdot
d\vec{S},\quad\quad\quad\quad\\
\dot{J}_{\mathrm tot}=\int_{\rm S}\left(r\rho v_\varphi\vec{\rm v}_{\rm p}-\frac{rB_\varphi\vec{B}_{\rm p}}{4\pi}\right)d\vec{S},
\label{mjdot}
\end{eqnarray}
where $\vec{S}$ is the surface of integration and $v_{\rm p}$ the poloidal velocity. We compute the values at half radius of the computation box, 25R$_\star$.

A typical distribution of jet velocities and the mass flux during the quasi-stationary state in our simulation is shown in Fig.~\ref{jet250}. Velocities and mass fluxes through the outflows above and below the disk midplane are different, as shown in Fig.~\ref{figb}. In both outflows, the mass flux is of the order of few per cents of the disk accretion rate. Velocities from the quasi-stationary interval are averaged for comparisons in the parameter study, which we table below. In Fig.~\ref{figc} we show the velocities and mass flux in the same case, exhibiting different values in the northern and southern outflows.

We tabulate our results in the parameter space with different magnetic fields strengths, disk resistivities $\alpha_{\rm m}$ and stellar rotational rates in units of Keplerian angular velocity at the equator, $\Omega_{\star} / \Omega_{\rm br}$. In Table~\ref{geom} are listed the obtained geometries, and in the Tables~\ref{mdottab} and \ref{jdottab} the obtained mass and angular momentum fluxes in the axial outflows above and below the disk mid-plane. In all the cases the viscosity coefficient is $\alpha_{\rm v}=1$.

In \cite{cem19} we found three possibilities DCE, DC, D for (D)isk+(C)olumn+(E)jection to which now the (A)xial outflow is added, as shown in Fig.~\ref{troj} and Table~\ref{geom}.

\begin{table}
\caption{Parameters in our simulations. With $\alpha_{\rm v}=1$, the corresponding magnetic Prandtl number P$_{\mathrm m}=\frac{2}{3}\alpha_{\mathrm v}/\alpha_{\mathrm m}$ for $\alpha_{\rm m}=$ 0.1, 0.4, 0.7, 1 is 6.7, 1.67, 0.95, 0.67, respectively. The simulation showcased in Figs.~\ref{jet250}-\ref{figc} is highlighted with a box and boldface letters, the letter code is explained in Section~\ref{res}. In the cases denoted with ``(-)'' simulations did not reach the quasi-stationary state. }
\centering                          
\begin{tabular}{ c c c c c }        
\hline    
\hline
$\alpha_{\rm m}=$ & 0.1 & 0.4  & 0.7  & 1\\ 
\hline\hline
 $\Omega_\star/\Omega_{\rm br}$ & & & & \\
\hline\hline
\multicolumn{5}{c}{$B_\star$=250~G} \\
0.5 & (DCEA) & \boxed{\bf{(DCA)}} & (DCA) & (DA) \\
0.8 & (-) & (DCA) & (DCA) & (DA) \\
\hline\hline     
\multicolumn{5}{c}{$B_\star$=500~G} \\
0.2 & (-) & (DCA) & (DCA) & (DCA) \\
0.5 & (DCEA) & (DCA) & (DCA) & (DCA) \\
0.8 & (DCEA) & (DA) & (DCA) & (DA) \\
\hline\hline    
\multicolumn{5}{c}{$B_\star$=750~G} \\
0.2 & (-) & (DCA) & (DCA) & (DCA) \\
0.5 & (DCEA) & (DCA) & (DCA) & (DCA) \\
0.8 & (DCEA) & (DCEA) & (DA) & (DA) \\
\hline\hline    
\multicolumn{5}{c}{$B_\star$=1000~G} \\
0.5 & (DCEA) & (DCA) & (DCA) & (DA) \\
0.8 & (DCEA) & (DCA) & (DA) & (DA) \\
\hline\hline    
\end{tabular}
\label{geom}
\end{table}
The mass fluxes are computed by using the fact that the system is axially symmetric, so we can perform the integral over the complete azimuthal angle $\varphi$:
\begin{equation}
\dot{M}=-2\pi R^2\int^\pi_0\rho(R,\theta)v_{\rm R}(R,\theta)\sin{\theta}d\theta .
\label{mdotss}
\end{equation}
The innermost part of the disk is very dense, and moves slowly towards the star, so that $v_{\rm R}<0$ and the overall contribution to $\dot{M}$ coming from the disk will be positive. Along the axis, above and below the disk midplane, launched are the axial outflows, which we assign as Northern and Southern Axial  Outflow (NAO and SAO), respectively. The material between the axial outflows in both directions is rarefied and contributes negligibly to the total mass flux, so we can write:
\begin{equation}
\dot{M}\approx\dot{M}_{\rm disk}-\dot{M}_{\rm NAO}-\dot{M}_{\rm SAO},
\label{summdot}
\end{equation}
where the $\dot{M}_{\rm NAO}$ and $\dot{M}_{\rm SAO}$ are integrated in their corresponding ranges of $\theta$, which is different in each case. By evaluating $\dot{M}_{\rm NAO}$ and $\dot{M}_{\rm SAO}$ and comparing them to $\dot{M}_{\rm disk}$, we quantify the magnitude of the outflows. If $\dot{M}_{\rm NAO}$ significantly differs from $\dot{M}_{\rm SAO}$, we consider the axial outflows asymmetric.

\begin{table}
\caption{Average value of the outflow rates $\dot{M}_{\rm NAO}/\dot{M}_{\rm SAO}$ in units of $10^{-10}~M_\odot$/y for the choices of the parameters as in Table~\ref{geom}. In each case the corresponding average accretion rate $\dot{M}$ in the same units is given in parentheses.} 
\centering                          
\begin{tabular}{ c c c c c }        
\hline    
\hline
$\alpha_{\rm m}=$ & 0.1 & 0.4  & 0.7  & 1\\ 
\hline\hline
 $\Omega_\star/\Omega_{\rm br}$ & & & & \\
\hline\hline
\multicolumn{5}{c}{$B_\star$=250~G} \\
0.5 & 3/3(10) & 10/11(15) & 9/7(11) & 2/5(2) \\
0.8 & - & 5/10(14) & 6/10(11) & 6/3(11) \\
\hline\hline     
\multicolumn{5}{c}{$B_\star$=500~G} \\
0.2 & - & 6/3(10) & 9/10(12) & 9/17(10) \\
0.5 & 1/9(7) & 8/7(12) & 6/7(13) & 1/3(12) \\
0.8 & 3/8(9) & 1/11(10) & 7/9(9) & 3/2(12) \\
\hline\hline    
\multicolumn{5}{c}{$B_\star$=750~G} \\
0.2 & 3/9(11) & 15/13(8) & 9/10(6) & 21/23(16) \\
0.5 & 2/7(8) & 25/21(20) & 6/9(13) & 10/13(10) \\
0.8 & 4/12(8) & 12/41(17) & 9/12(11) & 7/11(11) \\
\hline\hline    
\multicolumn{5}{c}{$B_\star$=1000~G} \\
0.5 & 1/7(9) & 23/18(16) & 13/1(7) & 17/8(20) \\
0.8 & 1/12(9) & 17/41(16) & 8/12(9) & 16/34(4) \\
\hline\hline    
\end{tabular}
\label{mdottab}
\end{table}
With the angular momentum flux we follow the same reasoning as with the mass flux. For the axisymmetric flow we compute it as:
\begin{equation}
\dot{J}=-2\pi R^3\int^\pi_0\left(\rho v_{\rm R}v_\varphi-\frac{B_{\rm R}B_\varphi}{4\pi}\right)\sin^2{\theta}d\theta .
\label{sumjdot}
\end{equation}
The integrals are computed at R=12R$_\star$.

\begin{table}
\caption{Average value of $\dot{J}_{\rm NAO}/\dot{J}_{\rm SAO}$ in units of J$_{\star 0}=\rho_{\rm d 0}R^4_\star v_{{\rm K}\star}$ for the choices of the parameters as in Table~\ref{geom}. In each case, the corresponding average accretion torque, $\dot{J}$, in the same units, is given in parentheses.}
\centering                          
\begin{tabular}{ c c c c c }        
\hline    
\hline
$\alpha_{\rm m}=$ & 0.1 & 0.4  & 0.7  & 1\\ 
\hline\hline
 $\Omega_\star/\Omega_{\rm br}$ & & & & \\
\hline\hline
\multicolumn{5}{c}{$B_\star$=250~G} \\
0.5 & 4/3(8) & 8/8(11) & 8/7(9) & 1/0.2(6) \\
0.8 & - & 3/4(7) & 3/7(6) & 4/2(6) \\
\hline\hline     
\multicolumn{5}{c}{$B_\star$=500~G} \\
0.2 & - &7/3(21) & 21/21(21) & 23/27(28) \\
0.5 & 1/8(6) & 9/15(9) & 10/7(21) & 6/10(9) \\
0.8 & 3/10(3) & 8/6(7) & 3/8(5) & 2/7(6) \\
\hline\hline    
\multicolumn{5}{c}{$B_\star$=750~G} \\
0.2 & - & 29/40(21) & 32/43(26) & 31/43 (28) \\
0.5 & 41/8(4) & 8/1(8) & 9/3(10) & 6/1(8) \\
0.8 & 8/3(3) & 2/7(4) & 2/8(5) & 2/2(6) \\
\hline\hline    
\multicolumn{5}{c}{$B_\star$=1000~G} \\
0.5 & 2/14(6) & 18/2(10) & 14/1(4) & 4/5(15) \\
0.8 & 3/16(3) & 10/43(15) & 5/6(4) & 4/16(4) \\
\hline\hline    
\end{tabular}
\label{jdottab}
\end{table}
%

\section{Conclusions}\label{con}
Our simulated outflows show that magnetospheric star-disk interaction is capable of launching axial outflows, which may form the core of observed jets. We show the results of a parameter study in which is determined the parameter space in our viscous and resistive MHD simulations in which axial outflows are launched. Our results corroborate the conclusions from \cite{R09}.

The presented simulations are an extension of the results from \citep{cem19} to the part of the parameter space where axial outflow is launched above each of the stellar poles. The obtained axial outflows are asymmetric for all choices of parameters in our simulations. For small values of $\alpha_{\rm m}\sim 0.1$ we typically obtain magnetospheric ejections, whereas for larger values like $\alpha_{\rm m}\sim 1$ the disk is pushed away from the star. Ejection in a zone above an accretion column occurs only at low values of $\alpha_{\rm m}$. Higher values of this parameter quench ejection and, if the stellar rotation rate is high enough, may even quench accretion by pushing the inner disk  away from the star. The strength of the stellar dipole field has a very modest influence on the accretion geometry.

The obtained magnetic torques have negative values of the rate of change of the angular momentum $\dot{J}$, resulting in the spin down of the star. We find that with the larger stellar rotation
rate, the stellar spin-down is larger. We also find that the axial outflows have a dominant influence on the rate of change of angular momentum and that the disc contribution has the same sign.

Our work here is a step between the parameter study with 2D axisymmetric simulations with the disk midplane taken as a boundary condition and the full 3D simulations. To obtain a stable axial outflow simulations were run for a large number of stellar rotations, in some cases more than one thousand. This imposed a change in the computational domain in comparison with \cite{cem19}, which complicates direct comparison of the results, but the trends in the results should remain preserved.

\section*{Acknowledgements}
This project was funded by a Polish NCN grant no. 2019/33/B/ST9/01564, and M\v{C} also acknowledges the Czech Science Foundation (GA\v{C}R) grant No.~21-06825X and the support by the International Space Science Institute (ISSI) in Bern, which hosted the International Team project \#495 (Feeding the spinning top) with its inspiring discussions. M\v{C} developed the PLUTO setup under ANR Toupies funding in CEA Saclay, France. We thank A. Mignone and his team of contributors for the possibility to use the PLUTO code, and ASIAA and CAMK for time on their Linux clusters XL and CHUCK, respectively.

\bibliographystyle{aa}
\bibliography{biblasimjets}

\end{document}